\documentstyle[12pt,epsfig]{article}
\newcommand{\agt}{\,\rlap{\lower 3.5 pt \hbox{$\mathchar \sim$}} \raise 1pt
 \hbox {$>$}\,}
\newcommand{\alt}{\,\rlap{\lower 3.5 pt \hbox{$\mathchar \sim$}} \raise 1pt
 \hbox {$<$}\,}
\catcode`@=11
\newcount\@tempcntc
\def\@citex[#1]#2{\if@filesw\immediate\write\@auxout{\string\citation{#2}}\fi
  \@tempcnta\z@\@tempcntb\m@ne\def\@citea{}\@cite{\@for\@citeb:=#2\do
    {\@ifundefined
       {b@\@citeb}{\@citeo\@tempcntb\m@ne\@citea\def\@citea{,}{\bf ?}\@warning
       {Citation `\@citeb' on page \thepage \space undefined}}%
    {\setbox\z@\hbox{\global\@tempcntc0\csname b@\@citeb\endcsname\relax}%
     \ifnum\@tempcntc=\z@ \@citeo\@tempcntb\m@ne
       \@citea\def\@citea{,}\hbox{\csname b@\@citeb\endcsname}%
     \else
      \advance\@tempcntb\@ne
      \ifnum\@tempcntb=\@tempcntc
      \else\advance\@tempcntb\m@ne\@citeo
      \@tempcnta\@tempcntc\@tempcntb\@tempcntc\fi\fi}}\@citeo}{#1}}
\def\@citeo{\ifnum\@tempcnta>\@tempcntb\else\@citea\def\@citea{,}%
  \ifnum\@tempcnta=\@tempcntb\the\@tempcnta\else
   {\advance\@tempcnta\@ne\ifnum\@tempcnta=\@tempcntb \else \def\@citea{--}\fi
    \advance\@tempcnta\m@ne\the\@tempcnta\@citea\the\@tempcntb}\fi\fi}
\catcode`@=12
\begin{document}
\title{\vskip-3cm{\baselineskip14pt
\centerline{\normalsize DESY~05--054\hfill ISSN~0418--9833}
\centerline{\normalsize INT--ACK--05--34\hfill}
\centerline{\normalsize hep--ph/0504058\hfill}
\centerline{\normalsize April 2005\hfill}}
\vskip1.5cm
$D^0$, $D^+$, $D_s^+$, and $\Lambda_c^+$ Fragmentation Functions from CERN
LEP1}
\author{Bernd A. Kniehl$^a$\thanks{Permanent address:
II. Institut f\"ur Theoretische Physik, Universit\"at Hamburg,
Luruper Chaussee 149, 22761 Hamburg, Germany.}
and Gustav Kramer$^b$\\
$^a$ Institute for Nuclear Theory, University of Washington,\\
Box 351550, Seattle, WA 98195-1550, USA\\
$^b$ II. Institut f\"ur Theoretische Physik, Universit\"at Hamburg,\\
Luruper Chaussee 149, 22761 Hamburg, Germany}
\date{}
\maketitle
\begin{abstract}
We present new sets of nonperturbative fragmentation functions for $D^0$,
$D^+$, and $D_s^+$ mesons as well as for $\Lambda_c^+$ baryons, both at leading
and next-to-leading order in the $\overline{\mathrm{MS}}$ factorization scheme
with five massless quark flavors.
They are determined by fitting data of $e^+e^-$ annihilation taken by the OPAL
Collaboration at CERN LEP1.
We take the charm-quark fragmentation function to be of the form proposed by
Peterson {\it et al.}\ and thus obtain new values of the $\epsilon_c$
parameter, which are specific for our choice of factorization scheme.

\end{abstract}
\newpage

\section{Introduction}

Several experimental collaborations at $ep$ and $p\overline{p}$ colliders
presented data on the differential cross section $d^2\sigma/dy\,dp_T$ for the
inclusive production of $D^0$, $D^+$, and $D_s^+$ mesons, $\Lambda_c^+$
baryons, and their charge-conjugate counterparts.
At DESY HERA, such data were collected by the ZEUS Collaboration
\cite{zeus,pad} in low-$Q^2$ $ep$ collisions, equivalent to photoproduction,
and by the H1 Collaboration \cite{h1} in deep-inelastic $ep$ scattering.
At the Fermilab Tevatron, such data were taken by the CDFII Collaboration
\cite{cdf} in $p\overline{p}$ collisions.

On the theoretical side, fragmentation functions (FF's) for the transitions
$c,b\to X_c$, where $X_c$ denotes a generic charmed hadron, are needed as
nonperturbative inputs for the calculation of all the cross sections mentioned
above.
Such FF's are preferably constructed by using precise information from
$e^+e^-\to X_c+X$ via $e^+e^-$ annihilation at the $Z$-boson resonance, where 
$X$ denotes the hadronic rest.
In this process, two mechanisms contribute with similar rates:
(i) $Z\to c\overline{c}$ decay followed by $c\to X_c$ (or
$\overline{c}\to X_c$) fragmentation; and
(ii) $Z\to b\overline{b}$ decay followed by $b\to X_b$ (or
$\overline{b}\to X_b$) fragmentation and weak $X_b\to X_c+X$ decay of the
bottom-flavored hadron $X_b$.
The latter two-step process is usually treated as a one-step fragmentation
process $b\to X_c$.

Using ALEPH \cite{aleph} and OPAL \cite{opal} data on inclusive $D^{*+}$
production at the $Z$-boson resonance, we determined separate FF's for
$c\to D^{*+}$ and $b\to D^{*+}$ in collaboration with Binnewies \cite{bkk}.
It is the purpose of this work to extract nonperturbative FF's for
$c,b\to D^0,D^+,D_s^+,\Lambda_c^+$ from the respective data samples collected
by the OPAL Collaboration at LEP1 \cite{opal1} using the same theoretical
framework as in Ref.~\cite{bkk}.

The work in Ref.~\cite{bkk} is based on the QCD-improved parton model
implemented in the modified minimal-subtraction ($\overline{\mathrm{MS}}$)
renormalization and factorization scheme in its pure form with $n_f=5$
massless quark flavors, which is also known the as the massless scheme
\cite{spira} or zero-mass variable-flavor-number scheme.
In this scheme, the masses $m_c$ and $m_b$ of the charm and bottom quarks are
neglected, except in the initial conditions of their FF's.
This is a reasonable approximation for center-of-mass (c.m.) energies
$\sqrt s\gg m_c,m_b$ in $e^+e^-$ annihilation or transverse momenta
$p_T\gg m_c,m_b$ in $ep$ and $p\overline{p}$ scattering, if the respective
FF's are used as inputs for the calculation of the cross sections for these
reactions.
Hence, we describe the $c,b\to X_c$ transitions by nonperturbative FF's, as is
usually done for the fragmentation of the up, down, and strange quarks into
light hadrons.

The outline of this paper is as follows.
In Sec.~\ref{sec:two}, we briefly recall the theoretical framework underlying
the extraction of FF's from the $e^+e^-$ data, which has already been
introduced in Refs.~\cite{bkk,bkk1}.
In Sec.~\ref{sec:three}, we present the $D^0$, $D^+$, $D_s^+$, and
$\Lambda_c^+$ FF's we obtained by fitting the respective LEP1 data samples
from OPAL \cite{opal1} at leading order (LO) and next-to-leading order (NLO)
in the massless scheme and discuss their properties.
In Sec.~\ref{sec:four}, we present predictions for the inclusive production of
these $X_c$ hadrons in nonresonant $e^+e^-$ annihilation at lower c.m.\
energies and compare them with data from other experiments.
Our conclusions are summarized in Sec.~\ref{sec:five}.

\section{Theoretical Framework}
\label{sec:two}

Our procedure to construct LO and NLO sets of $D$ FF's has already been
described in Refs.~\cite{bkk,bkk1}.
As experimental input, we use the LEP1 data from OPAL \cite{opal1}.

In $e^+e^-$ annihilation at the $Z$-boson resonance, $X_c$ hadrons are
produced either directly through the hadronization of charm quarks produced 
by $Z\to c\overline{c}$ or via the weak decays of $X_b$ hadrons from
$Z\to b\overline{b}$.
In order to disentangle these two production modes, the authors of
Ref.~\cite{opal1} utilized the apparent decay length distributions and energy
spectra of the $X_c$ hadrons.
Because of the relatively long $X_b$-hadron lifetimes and the hard $b\to X_b$
fragmentation, $X_c$ hadrons originating from $X_b$-hadron decays have
significantly longer apparent decay lengths than those from primary production.
In addition, the energy spectrum of $X_c$ hadrons originating from
$X_b$-hadron decays is much softer than that due to primary charm production. 

The experimental cross sections \cite{opal1} were presented as distributions
differential in $x=2E(X_c)/\sqrt s$, where $E(X_c)$ is the measured energy of
the $X_c$-hadron candidate, and normalized to the total number of hadronic
$Z$-boson decays.
Besides the total $X_c$ yield, which receives contributions from $Z\to c\bar c$
and $Z\to b\bar b$ decays as well as from light-quark and gluon fragmentation,
the OPAL Collaboration separately specified results for $X_c$ hadrons from
tagged $Z\to b\bar b$ events.
As already mentioned above, the contribution due to charm-quark fragmentation
is peaked at large $x$, whereas the one due to bottom-quark fragmentation has
its maximum at small $x$.

For the fits, we use the $x$ bins in the interval $[0.15,1.0]$ and integrate
the theoretical cross sections over the bin widths used in the experimental
analysis.
For each of the four charmed-hadron species considered here,
$X_c=D^0,D^+,D_s^+,\Lambda_c^+$, we sum over the two charge-conjugate states as
was done in Ref.~\cite{opal1}.
As a consequence, there is no difference between the FF's of a given quark
and its antiquark.
As in Refs.~\cite{bkk,bkk1}, we take the starting scales for the $X_c$ FF's of
the gluon and the $u$, $d$, $s$, and $c$ quarks and antiquarks to be
$\mu_0=2m_c$, while we take $\mu_0=2m_b$ for the FF's of the bottom quark and
antiquark.
The FF's of the gluon and the first three flavors are assumed to be zero at
their starting scale.
At larger scales $\mu$, these FF's are generated through the usual
Dokshitzer-Gribov-Lipatov-Altarelli-Parisi (DGLAP) \cite{dglap} evolution at LO
or NLO.
The FF's of the first three quarks and antiquarks coincide with each other at
all scales $\mu$.

We employ two different forms for the parameterizations of the charm- and
bottom-quark FF's at their respective starting scales.
In the case of charm, we use the distribution of Peterson {\it et al.}\
\cite{pet},
\begin{equation}
D_c(x,\mu_0^2)=N\frac{x(1-x)^2}{[(1-x)^2+\epsilon x]^2}.
\label{eq:peterson}
\end{equation}
In the case of bottom, we adopt the ansatz
\begin{equation}
D_b(x,\mu_0^2)=Nx^{\alpha}(1-x)^{\beta},
\label{eq:standard}
\end{equation}
which is frequently used for the FF's of light hadrons.
Equation~(\ref{eq:peterson}) is particularly suitable for FF's that peak at
large values of $x$, as is typically the case for $c\to X_c$ transitions.
Since the $b\to X_c$ FF is a convolution of the $b\to X_b$ fragmentation and
the subsequent $X_b\to X_c+X$ decay, it has its maximum at small $x$ values.
Therefore, Eq.~(\ref{eq:peterson}) is less suitable in this case.
We apply Eqs.~(\ref{eq:peterson}) and (\ref{eq:standard}) for the FF's of all 
four $X_c$-hadron species considered here.

The calculation of the cross section $(1/\sigma_{\rm tot})d\sigma/dx$ for
$e^+e^-\to\gamma/Z\to X_c+X$ is performed as described in Ref.~\cite{bkk}, in
the pure $\overline{\mathrm{MS}}$ subtraction scheme, {\it i.e.}, without the
subtraction terms $d_{Qa}(x)$ specified in Eq.~(2) of Ref.~\cite{kks}.
All relevant formulas and references may be found in Ref.~\cite{bkk1}.
As for the asymptotic scale parameter for five active quark flavors, we adopt
the LO (NLO) value $\Lambda_{\overline{\rm MS}}^{(5)}=108$~MeV (227~MeV) from
our study of inclusive charged-pion and -kaon production \cite{bkk2}.
The particular choice of $\Lambda_{\overline{\rm MS}}^{(5)}$ is not essential,
since other values can easily accommodated by slight shifts of the other fit
parameters.
As in Refs.~\cite{bkk,bkk1}, we take the charm- and bottom-quark masses to be  
$m_c=1.5$~GeV and $m_b=5$~GeV, respectively.

\boldmath
\section{Determination of the $D^0$, $D^+$, $D_s^+$, and $\Lambda_c^+$ FF's}
\label{sec:three}
\unboldmath

The OPAL Collaboration \cite{opal1} presented $x$ distributions for their full
$D^0$, $D^+$, $D_s^+$, and $\Lambda_c^+$ samples and for their $Z\to b\bar b$
subsamples.
We received these data in numerical form via private communication
\cite{martin}.
They are displayed in Figs.~4 (for the $D^0$ and $D^+$ mesons) and 5 (for the
$D_s^+$ meson and the $\Lambda_c^+$ baryon) of Ref.~\cite{opal1} in the form
$(1/N_{\rm had})dN/dx$, where $N$ is the number of $X_c$-hadron candidates
reconstructed through appropriate decay chains.
In order to convert this into the cross sections
$(1/\sigma_{\rm tot})d\sigma/dx$, we need to divide by the branching 
fractions of the decays that were used in Ref.~\cite{opal1} for the
reconstruction of the various $X_c$ hadrons, namely,
\begin{eqnarray}
B(D^0\to K^-\pi^+)&=&(3.84\pm0.13)\%,\nonumber\\
B(D^+\to K^-\pi^+\pi^-)&=&(9.1\pm0.6)\%,\nonumber\\
B\left(D_s^+\to \phi\pi^+\right)&=&(3.5\pm0.4)\%,\nonumber\\
B\left(\Lambda_c^+\to pK^-\pi^+\right)&=&(4.4\pm0.6)\%,
\end{eqnarray}
respectively.
The experimental errors on these branching fractions are not included in our
analysis.

The values of $N$ and $\epsilon$ in Eq.~(\ref{eq:peterson}) and of $N$,
$\alpha$, and $\beta$ in Eq.~(\ref{eq:standard}) which result from our LO and
NLO fits to the OPAL data are collected in Table~\ref{tab:par}.
From there, we observe that the parameters $\alpha$ and $\beta$, which
characterize the shape of the bottom FF, take very similar values for the
various $X_c$ hadrons, which are also similar to those for the $D^{*+}$
meson listed in Table~I of Ref.~\cite{bkk}.
On the other hand, the values of the $\epsilon$ parameter, which determines
the shape of the charm FF, significantly differ from particle species to
particle species.
In the $D^{*+}$ case \cite{bkk}, our LO (NLO) fits to ALEPH \cite{aleph} and
OPAL \cite{opal} data, which required separate analyses, yielded
$\epsilon=0.144$ (0.185) and 0.0851 (0.116), respectively.
We observe that, for each of the $X_c$-hadron species considered, the LO
results for $\epsilon$ are considerably smaller than the NLO ones.
Furthermore, we notice a tendency for the value of $\epsilon$ to decrease as
the mass ($m_{X_c}$) of the $X_c$ hadron increases. 

\begin{table}
\begin{center}
\caption{Fit parameters of the charm- and bottom-quark FF's for the various
$X_c$ hadrons at LO and NLO.
The corresponding starting scales are $\mu_0=2m_c=3$~GeV and 
$\mu_0=2m_b=10$~GeV, respectively.
All other FF's are taken to be zero at $\mu_0=2m_c$.}
\label{tab:par}
\begin{tabular}{ccccccc}
\hline\hline
$X_c$ & Order & $Q$ & $N$ & $\alpha$ & $\beta$ & $\epsilon$ \\
\hline
$D^0$         & LO  & $c$ & 0.998  & --   & --   & 0.163  \\
              &     & $b$ & 71.8   & 1.65 & 5.19 & --     \\
              & NLO & $c$ & 1.16   & --   & --   & 0.203  \\
              &     & $b$ & 97.5   & 1.71 & 5.88 & --     \\
$D^+$         & LO  & $c$ & 0.340  & --   & --   & 0.148  \\
              &     & $b$ & 48.5   & 2.16 & 5.38 & --     \\
              & NLO & $c$ & 0.398  & --   & --   & 0.187  \\
              &     & $b$ & 64.9   & 2.20 & 6.04 & --     \\
$D_s^+$       & LO  & $c$ & 0.0704 & --   & --   & 0.0578 \\
              &     & $b$ & 40.0   & 2.05 & 4.93 & --     \\
              & NLO & $c$ & 0.0888 & --   & --   & 0.0854 \\
              &     & $b$ & 21.8   & 1.64 & 4.71 & --     \\
$\Lambda_c^+$ & LO  & $c$ & 0.0118 & --   & --   & 0.0115 \\
              &     & $b$ & 44.1   & 1.97 & 6.33 & --     \\
              & NLO & $c$ & 0.0175 & --   & --   & 0.0218 \\
              &     & $b$ & 27.3   & 1.66 & 6.24 & --     \\
\hline\hline
\end{tabular}
\end{center}
\end{table}

In Table~\ref{tab:chi}, we list three values of $\chi^2$ per degree of freedom
($\chi_{\rm DF}^2$) for each of the fits from Table~\ref{tab:par}:
one for the $Z\to b\overline{b}$ subsample, one for the total sample (sum of
tagged-$c\overline{c}$, tagged-$b\overline{b}$, and gluon-splitting events),
and an average one evaluated by taking into account the $Z\to b\overline{b}$
subsample and the total sample.
The actual $\chi_{\rm DF}^2$ values are rather small.
This is due to the sizeable errors and the rather limited number of data
points, especially for the $D_s^+$ and $\Lambda_c^+$ data.
In each case, the $Z\to b\overline{b}$ subsample is somewhat less well
described than the total sample.
The NLO fits yield smaller $\chi_{\rm DF}^2$ values than the LO ones, except
for the $\Lambda_c^+$ case.

\begin{table}
\begin{center}
\caption{$\chi^2$ per degree of freedom achieved in the LO and NLO fits to the
OPAL \cite{opal1} data on the various $D$ hadrons.
In each case, $\chi_{\rm DF}^2$ is calculated for the $Z\to b\overline{b}$
sample ($b$), the full sample (All), and the combination of both (Average).}
\label{tab:chi}
\begin{tabular}{ccccc}
\hline\hline
$X_c$ & Order & $b$ & All & Average \\
\hline
$D^0$         & LO  & 1.16  & 0.688 & 0.924 \\
              & NLO & 0.988 & 0.669 & 0.829 \\
$D^+$         & LO  & 0.787 & 0.540 & 0.663 \\
              & NLO & 0.703 & 0.464 & 0.584 \\
$D_s^+$       & LO  & 0.434 & 0.111 & 0.273 \\
              & NLO & 0.348 & 0.108 & 0.228 \\
$\Lambda_c^+$ & LO  & 1.05  & 0.106 & 0.577 \\
              & NLO & 1.05  & 0.118 & 0.582 \\
\hline\hline
\end{tabular}
\end{center}
\end{table}

\begin{figure}[ht]
\begin{center}
\epsfig{figure=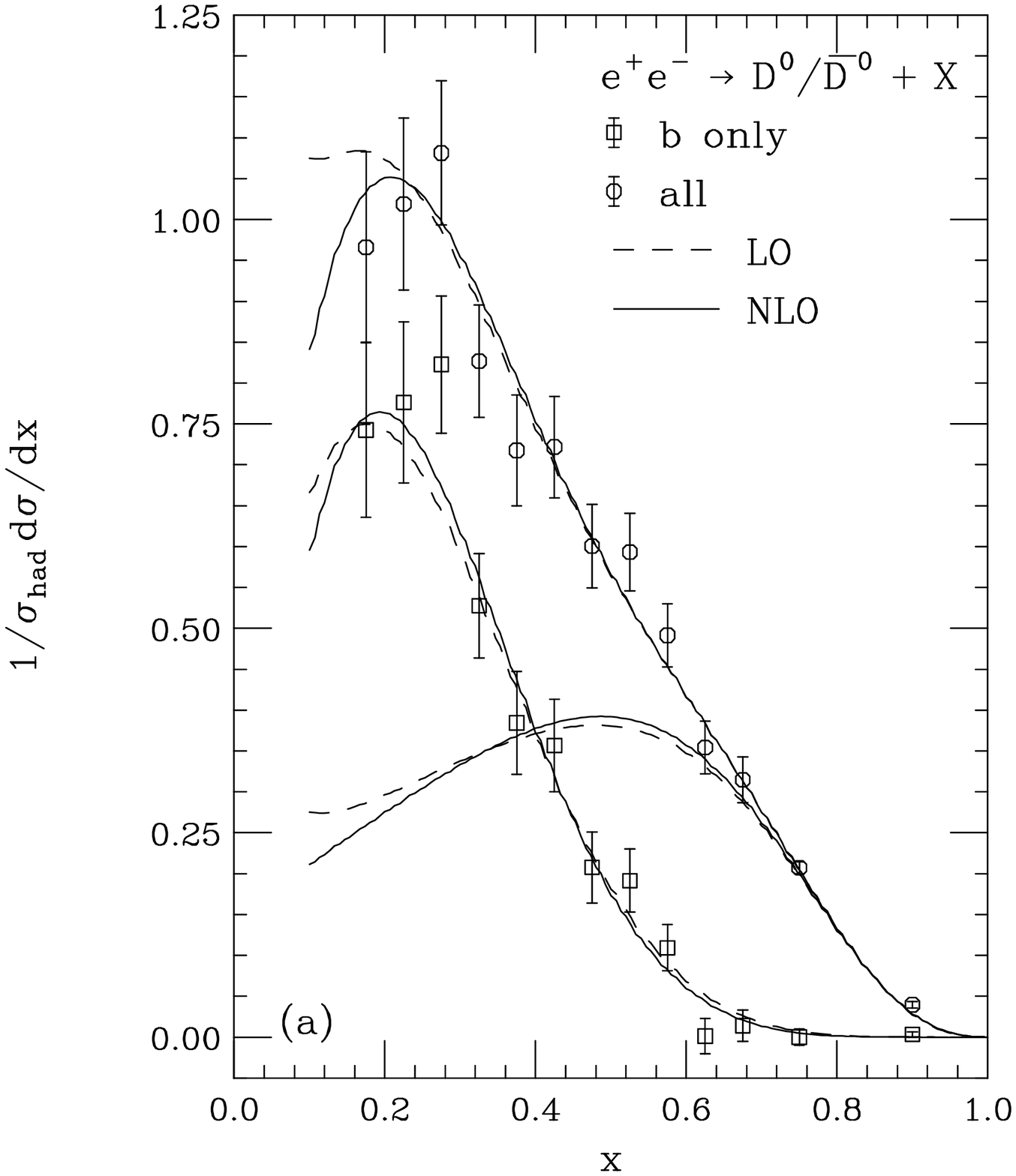,width=\textwidth}
\caption{The normalized differential cross sections
$(1/\sigma_{\rm tot})d\sigma/dx$ of inclusive (a) $D^0/\overline{D}^0$, (b)
$D^\pm$, (c) $D_s^\pm$, and (d) $\Lambda_c^\pm$ production in $e^+e^-$
annihilation on the $Z$-boson resonance evaluated at LO (dashed lines) and NLO
(solid lines) with our respective FF sets are compared with the OPAL data
\cite{opal1} renormalized as explained in the text (circles).
The same is also done for the $Z\to b\overline{b}$ subsamples (squares).
In addition, our LO and NLO fit results for the $Z\to c\overline{c}$
contributions are shown.
In each case, the $X_c$ hadron and its charge-conjugate counterpart are summed
over.}
\label{fig:xs}
\end{center}
\end{figure}

The normalized differential cross sections $(1/\sigma_{\rm tot})d\sigma/dx$
for $D^0$, $D^+$, $D_s^+$, and $\Lambda_c^+$ hadrons (circles), extracted from
Ref.~\cite{opal1} as explained above, are compared with our LO (upmost dashed
lines) and NLO (upmost solid lines) fits in Figs.~\ref{fig:xs}(a)--(d),
respectively.
The same is also done for the $Z\to b\overline{b}$ subsamples (squares).
In addition, our LO and NLO fit results for the $Z\to c\overline{c}$
contributions are shown.
In each case, the $X_c$ hadron and its charge-conjugate partner are summed
over.
From Figs.~\ref{fig:xs}(a)--(d), we observe that the LO and NLO results are
very similar, except for very small values of $x$.
This is also true for the distributions at the starting scales, as may be seen
by comparing the corresponding LO and NLO parameters in Table~\ref{tab:par}.
The branching of the LO and NLO results at small values of $x$ indicates that,
in this region, the perturbative treatment ceases to be valid. 
This is related to the phase-space boundary for the production of $X_c$ hadrons
at $x_{\rm min}=2m_{X_c}/\sqrt s$.
These values are somewhat larger than the $x$ values where our NLO results turn
negative.
Since our massless-quark approach is not expected to be valid in regions of
phase space where finite-$m_{X_c}$ effects are important, our results should
only be considered meaningful for $x\agt x_{\rm cut}=0.1$, say.
We also encountered a similar small-$x$ behavior for the $D^{*+}$ FF's in
Refs.~\cite{bkk,bkk1}.

As mentioned above, we take the FF's of the partons
$g,u,\overline{u},d,\overline{d},s,\overline{s}$ to be vanishing at their
starting scale $\mu_0=2m_c$.
However, these FF's are generated via the DGLAP evolution to the high scale
$\mu=\sqrt s$.
Thus, apart from the FF's of the heavy quarks $c,\overline{c},b,\overline{b}$,
also these radiatively generated FF's contribute to the cross section.
All these contributions are properly included in the total result for
$(1/\sigma_{\rm tot})d\sigma/dx$ shown in Figs.~\ref{fig:xs}(a)--(d).
At LEP1 energies, the contribution from the first three quark flavors is still
negligible; it is concentrated at small values of $x$ and only amounts to a
few percent of the integrated cross section.
However, the contribution from the gluon FF, which appears at NLO in 
connection with $q\overline{q}g$ final states, is numerically significant.
As in our previous works \cite{bkk,bkk1}, motivated by the decomposition of
$(1/\sigma_{\rm tot})d\sigma/dx$ in terms of parton-level cross sections, we
distributed this contribution over the $Z\to c\bar c$ and $Z\to b\bar b$
channels in the ratio $e_c^2:e_b^2$, where $e_q$ is the effective electroweak
coupling of the quark $q$ to the $Z$ boson and the photon including propagator
adjustments.
This procedure should approximately produce the quantities that are compared
with the OPAL data \cite{opal1}.

As in Refs.~\cite{bkk,bkk1}, we study the branching fractions for the
transitions\break $c,b\to D^0,D^+,D_s^+,\Lambda_c^+$, defined by
\begin{equation}
B_Q(\mu)=\int_{x_{\rm cut}}^1dx\,D_Q(x,\mu^2),
\label{eq:br}
\end{equation}
where $Q=c,b$, $D_Q$ are the appropriate FF's, and $x_{\rm cut}=0.1$.
This allows us to test the consistency of our fits with information presented
in the experimental paper \cite{opal1} that was used for our fits.
The contribution from the omitted region $0<x<x_{\rm cut}$ is small.
Table~\ref{tab:br} contains the values of $B_Q(\mu)$ for all eight transitions
$c,b\to D^0,D^+,D_s^+,\Lambda_c^+$ evaluated according to Eq.~(\ref{eq:br}) in
LO and NLO at the respective thresholds $\mu=2m_Q$ and at the $Z$-boson
resonance $\mu=M_Z$.
As expected, the values of $B_Q(\mu)$ change very little under the evolution
from $\mu=2m_Q$ to $\mu=M_Z$, and they are rather similar for $Q=c,b$.
Leaving aside the insignificant contribution due to strange charm baryons,
the values of $B_Q(\mu)$ for $X_c=D^0,D^+,D_s^+,\Lambda_c^+$ should
approximately add up to unity for each heavy flavor $Q=c,b$ at any value of
$\mu$.
Although we did not impose this sum rule as a constraint on our fits, it is
well satisfied for $B_c(M_Z)$ and $B_b(M_Z)$ at NLO.
In fact, from Table~\ref{tab:br} one obtains 103\% and 99.5\%, respectively.
The corresponding LO values, being 108\% and 105\%, are somewhat too large, as
may be understood by observing the excess of the LO fits over the experimental
data at small values of $x$ in Figs.~\ref{fig:xs}(a)--(d).
The corresponding sums of the LO and NLO entries for $B_c(2m_c)$ and
$B_b(2m_b)$ in Table~\ref{tab:br} range between 110\% and 116\%.
In view of the long evolution paths from the charm and bottom thresholds way
up to the $Z$-boson resonance, such violations of the sum rule can be
considered acceptable.
The situation is expected to improve once experimental data at lower c.m.\
energies (see Sec.~\ref{sec:four}) are included in our fits.

\begin{table}
\begin{center}
\caption{Branching fractions (in \%) of $c,b\to D^0,D^+,D_s^+,\Lambda_c^+$
evaluated according to Eq.~(\ref{eq:br}) in LO and NLO at the respective
starting scales $\mu=2m_Q$ and at the $Z$-boson resonance $\mu=M_Z$.}
\label{tab:br}
\begin{tabular}{cccccc}
\hline\hline
$X_c$ & Order & $B_c(2m_c)$ & $B_c(M_Z)$ & $B_b(2m_b)$ & $B_b(M_Z)$ \\
\hline
$D^0$         & LO  & 72.1  & 66.9  & 57.8 & 52.8 \\
              & NLO & 69.5  & 63.9  & 55.2 & 49.8 \\
$D^+$         & LO  & 26.6  & 24.7  & 19.4 & 17.9 \\
              & NLO & 25.6  & 23.6  & 18.6 & 17.1 \\
$D_s^+$       & LO  & 11.5  & 10.9  & 22.4 & 20.6 \\
              & NLO & 10.8  & 10.1  & 21.6 & 19.6 \\
$\Lambda_c^+$ & LO  &  5.88 &  5.67 & 15.1 & 13.7 \\
              & NLO &  5.74 &  5.48 & 14.5 & 13.0 \\
\hline\hline
\end{tabular}
\end{center}
\end{table}

It is interesting to compare our LO and NLO values of $B_c(M_Z)$ and $B_b(M_Z)$
for the $D^0$ ,$D^+$, $D_s^+$, and $\Lambda_c^+$ hadrons with the respective
results determined by the OPAL Collaboration through Peterson model fits.
These results are presented in Table~9 of Ref.~\cite{opal1} in the dressed form
\begin{equation}
p_{Q\to X_c}=R_QB_Q(M_Z)B_{X_c},
\label{eq:pr}
\end{equation}
where $R_Q=\Gamma_{Q\overline{Q}}/\Gamma_{\rm had}$ are the production rates of
the quarks $Q=c,b$ in $e^+e^-$ annihilation on the $Z$-boson resonance and
$B_{X_c}$ are the decay branching fractions of the four $X_c$ hadrons
$X_c=D^0,D^+,D_s^+,\Lambda_c^+$ into the channels considered in
Eq.~(\ref{eq:br}).
For the reader's convenience, these results are copied to Table~\ref{tab:pr},
where they are compared with our results for $p_{Q\to X_c}$, which are
obtained from the appropriate entries in Table~\ref{tab:br} through
multiplication with the branching fractions from Eq.~(\ref{eq:br}) and the
production rates $R_c=0.1689\pm0.0047$ and $R_b=0.21643\pm0.00072$ determined
by the Particle Data Group \cite{pdg} in the framework of the Standard Model.
For simplicity, the values deduced from Table~\ref{tab:pr} do not include the
errors on $R_Q$ and $B_{X_c}$ and those on $B_Q(M_Z)$ resulting from our fits.

In Table~9 of Ref.~\cite{opal1}, the OPAL Collaboration also presented the
total rates $\overline{n}(Z\to X_c)B_{X_c}$, which include the estimated
contributions from gluon splitting $g\to Q\overline{Q}$; for further details,
see Ref.~\cite{opal1}.
In Table~\ref{tab:pr}, these results are quoted and compared with the
corresponding quantities $2(p_{c\to X_c}+p_{b\to X_c})$ resulting from our LO
and NLO analyses.
Notice that the experimental results are corrected to include the unmeasured
contributions from $x<0.15$, whereas our evaluations of Eq.~(\ref{eq:br})
exclude the contributions from $x<x_{\rm cut}$.
This explains why the experimental results somewhat overshoot ours.
The agreement is worse at NLO, which may be understood by observing that our
evaluations of $2(p_{c\to X_c}+p_{b\to X_c})$ do not include the
contributions from gluon fragmentation, which enters the stage at NLO.
Keeping these caveats in mind, we find reasonable overall agreement between
the OPAL results for $\overline{n}(Z\to X_c)B_{X_c}$ and our results for
$2(p_{c\to X_c}+p_{b\to X_c})$.

{\scriptsize
\begin{table}
\begin{center}
\caption{$X_c$-hadron production rates reported by OPAL \cite{opal1} compared
to results evaluated at LO and NLO from Eq.~(\ref{eq:pr}) using the branching
fractions from Table~\ref{tab:br}.}
\label{tab:pr}
\begin{tabular}{ccccccc}
\hline\hline
$X_c$ & \multicolumn{2}{c}{$p_{c\to X_c}$ [\%]} &
\multicolumn{2}{c}{$p_{b\to X_c}$ [\%]} &
\multicolumn{2}{c}{$\overline{n}(Z\to X_c)B_{X_c}$ [\%]} \\
 & \cite{opal1} & fit ${{\rm LO}\atop{\rm NLO}}$ &
\cite{opal1} & fit ${{\rm LO}\atop{\rm NLO}}$ &
\cite{opal1} & fit ${{\rm LO}\atop{\rm NLO}}$ \\
\hline
$D^0$ & $0.389\pm0.027{+0.026\atop-0.024}$ & 0.434 &
$0.454\pm0.023{+0.025\atop-0.026}$ & 0.439 &
$1.784\pm0.066\pm0.086$ & 1.746 \\
 & & 0.414 & & 0.414 & & 1.656 \\
$D^+$ & $0.358\pm0.046{+0.025\atop-0.031}$ & 0.380 &
$0.379\pm0.031{+0.028\atop-0.025}$ & 0.353 &
$1.548\pm0.082{+0.082\atop-0.080}$ & 1.466 \\
 & & 0.363 & & 0.337 & & 1.400 \\
$D_s^+$ & $0.056\pm0.015\pm0.007$ & 0.0644 &
$0.166\pm0.018\pm0.016$ & 0.156 &
$0.460\pm0.036\pm0.040$ & 0.441 \\
 & & 0.0597 & & 0.148 & & 0.415 \\
$\Lambda_c^+$ & $0.041\pm0.019\pm0.007$ & 0.0421 &
$0.122\pm0.023\pm0.010$ & 0.130 &
$0.345\pm0.052\pm0.029$ & 0.344 \\
 & & 0.0407 & & 0.124 & & 0.329 \\
\hline\hline
\end{tabular}
\end{center}
\end{table}
}

Our LO and NLO values of $B_c(M_Z)$ and $B_b(M_Z)$ for the $D^0$, $D^+$,
$D_s^+$, and $\Lambda_c^+$ hadrons in Table~\ref{tab:br} can also be compared
with experimental results published more recently by the ALEPH \cite{aleph}
and DELPHI \cite{delphi} Collaborations.
In Ref.~\cite{aleph}, $B_c(M_Z)$ are called $f(c\to X_c)$ and may be found in
Secs.~7.1 and 7.3.
In Ref.~\cite{delphi}, $B_Q(M_Z)$ are called $P_{Q\to X_c}$ and may be
extracted for $Q=c$ from Table 13 (in connection with sum rule of Eq.~(12) and
taking into account the discussion of the contribution from the strange charm
baryons in Sec.~8.2) and for $Q=b$ from Table 15.
For simplicity, we add the three types of errors quoted in
Refs.~\cite{aleph,delphi} (from statistics, systematics, and decay branching
fractions) in quadrature.
In 1999, Gladilin \cite{gla} derived world-average values of $B_c(M_Z)$ for
the $D^0$ ,$D^+$, $D_s^+$, and $\Lambda_c^+$ hadrons related to $e^+e^-$
annihilation, which are also listed in Table~\ref{tab:ad}.

The branching fractions of the $c\to D^0,D^+,D_s^+,\Lambda_c^+$ transitions
were also measured in $ep$ collisions at HERA, in photoproduction by the ZEUS
Collaboration \cite{zeus,pad} and in deep-inelastic scattering by the H1
Collaboration \cite{h1}.
These results are also included in Table~\ref{tab:ad} for comparison.
Strictly speaking, they do not correspond to $B_c(M_Z)$, but rather to
$B_c(\mu)$, where $\mu$ is set by the average value of $p_T$ (in the case of
photoproduction) or $Q$ (in the case of deep-inelastic scattering).
However, from Table~\ref{tab:br} we know that the $\mu$ dependence of
$B_c(\mu)$ is relatively mild.

We observe that the experimental results collected in Table~\ref{tab:ad}, which
are mostly independent from each other, are mutually consistent within errors.
Comparing them with the corresponding entries in the forth and sixth columns
of Table~\ref{tab:br}, we find resonable overall agreement. 

\begin{table}
\begin{center}
\caption{Branching fractions (in \%) of $c,b\to D^0,D^+,D_s^+,\Lambda_c^+$
reported by ALEPH \cite{aleph}, DELPHI \cite{delphi}, Gladilin \cite{gla},
ZEUS \cite{pad}, and H1 \cite{h1}.}
\label{tab:ad}
\begin{tabular}{ccccccc}
\hline\hline
$X_c$ & \multicolumn{5}{c}{$B_c(M_Z)$}& $B_b(M_Z)$ \\
 & \cite{aleph} & \cite{delphi} & \cite{gla} & \cite{pad} & \cite{h1} &
\cite{delphi} \\
\hline
$D^0$ & $55.9\pm2.2$ & $54.80\pm4.78$ & $54.9\pm2.6$ &
$55.7{+2.0\atop-2.3}$ & $65.8{+15.5\atop-15.9}$ & $60.05\pm4.39$ \\ 
$D^+$ & $23.79\pm2.42$ & $22.70\pm1.82$ & $23.2\pm1.8$ &
$24.9{+1.5\atop-1.6}$ & $20.2{+5.7\atop-4.4}$ & $23.01\pm2.13$ \\ 
$D_s^+$ & $11.6\pm3.6$ & $12.51\pm2.97$ & $10.1\pm2.7$ &
$10.7\pm1.0$ & $15.6{+7.5\atop-7.2}$ & $16.65\pm4.50$ \\ 
$\Lambda_c^+$ & $7.9\pm2.2$ & $8.76\pm3.30$ & $7.6\pm2.1$ &
$7.6{+2.6\atop-2.0}$ & -- & $8.90\pm3.00$ \\
\hline\hline
\end{tabular}
\end{center}
\end{table}

Another quantity of interest, which can directly be compared with experiment,
is the mean momentum fraction,
\begin{equation}
\langle x\rangle_Q(\mu)=\frac{1}{B_Q(\mu)}\int_{x_{\rm cut}}^1dx\,xD_Q(x,\mu).
\end{equation}
In Table~\ref{tab:xav}, we present the values of $\langle x\rangle_Q(\mu)$ for
$Q=c,b$ evaluated at $\mu=2m_Q,M_Z$ with the LO and NLO FF's of the $D^0$,
$D^+$, $D_s^+$, and $\Lambda_c^+$ hadrons.
At fixed value of $\mu$, the differences between the LO and NLO sets are
insignificant.
The DGLAP evolution from $\mu=2m_Q$ to $\mu=M_Z$ leads to a significant 
reduction of $\langle x\rangle_Q(\mu)$, especially in the case of $Q=c$.
The values of $\langle x\rangle_b(\mu)$ are appreciably smaller than the
values of $\langle x\rangle_c(\mu)$, as is expected because the bottom-quark
fragmentation into $X_c$ hadrons is much softer than the charm-quark one.

Our values of $\langle x\rangle_c(M_Z)$ for the $D^0$ and $D^+$ mesons should
be compared with the respective results obtained by the OPAL Collaboration
\cite{opal1} in the framework of the Peterson model \cite{pet}, which read
\begin{eqnarray}
\langle x\rangle_c(M_Z)&=&0.487\pm0.009{+0.011\atop-0.009}
\qquad(D^0),
\nonumber\\
\langle x\rangle_c(M_Z)&=&0.483\pm0.015{+0.007\atop-0.011}
\qquad(D^+)
\label{eq:xav}
\end{eqnarray}
for the $D^0$ and $D^+$ mesons, respectively.
The differences to the values obtained for three other fragmentation models
are included in the systematical errors.
Comparing Eq.~(\ref{eq:xav}) with the corresponding entries in
Table~\ref{tab:xav}, we observe that the latter are slightly smaller.

\begin{table}
\begin{center}
\caption{Average momentum fractions of $c,b\to D^0,D^+,D_s^+,\Lambda_c^+$
evaluated according to Eq.~(\ref{eq:xav}) in LO and NLO at the respective
starting scales $\mu=2m_Q$ and at the $Z$-boson resonance $\mu=M_Z$.}
\label{tab:xav}
\begin{tabular}{cccccc}
\hline\hline
$X_c$ & Order & $\langle x\rangle_c(2m_c)$ & $\langle x\rangle_c(M_Z)$ &
$\langle x\rangle_b(2m_b)$ & $\langle x\rangle_b(M_Z)$ \\
\hline
$D^0$         & LO  & 0.588 & 0.452 & 0.316 & 0.284 \\
              & NLO & 0.568 & 0.431 & 0.300 & 0.270 \\
$D^+$         & LO  & 0.596 & 0.458 & 0.341 & 0.303 \\
              & NLO & 0.575 & 0.436 & 0.323 & 0.287 \\
$D_s^+$       & LO  & 0.676 & 0.512 & 0.349 & 0.310 \\
              & NLO & 0.644 & 0.482 & 0.332 & 0.296 \\
$\Lambda_c^+$ & LO  & 0.791 & 0.590 & 0.302 & 0.273 \\
              & NLO & 0.750 & 0.553 & 0.288 & 0.261 \\
\hline\hline
\end{tabular}
\end{center}
\end{table}

\boldmath
\section{Comparison with $e^+e^-$ data at lower energies}
\label{sec:four}
\unboldmath

The fractional energy spectra of inclusive $D^0$, $D^+$, $D_s^+$, and
$\Lambda_c^+$ production was also measured in nonresonant $e^+e^-$ annihilation
at lower energies.
Specifically, the CLEO Collaboration took $D^0$, $D^+$ (Table~XII in
Ref.~\cite{cleo04}), $D_s^+$ (Table~IV in Ref.~\cite{cleo0}), and
$\Lambda_c^+$ (Table~V in Ref.~\cite{cleo88}) data at LEPP CESR with
$\sqrt s=10.55$~GeV;
the HRS Collaboration took $D^0$, $D^+$ (Table~1 in Ref.~\cite{hrs88}), and
$D_s^+$ (Table~I in Ref.~\cite{hrs85}) data at SLAC PEP with $\sqrt s=29$~GeV;
and the TASSO Collaboration took $D_s^+$ (Fig.~3 in Ref.~\cite{tasso}) data
at DESY PETRA with $\sqrt s=34.7$~GeV.
It is instructing to confront these data with LO and NLO predictions based on
our new FF's, so as to test the scaling violations predicted by the DGLAP
evolution equations.
An especially interesting situation arises for the CLEO data
\cite{cleo04,cleo0,cleo88}, from which all $X_c$ hadrons coming from
$X_b$-hadron decays are excluded by appropriate acceptance cuts, so that only
$n_f=4$ quark flavors are active and a direct test of the charm-quark FF's is
feasible.

The $D^+$ and $\Lambda_c^+$ data explicitly refer to the decay channels
$D_s^+\to\phi\pi^+\to K^+K^-\pi^+$ and $\Lambda_c^+\to pK^-\pi^+$,
respectively, and we have to divide them by the corresponding branching
fractions.
For this, we use the up-to-date values
$B(D_s^+\to\phi\pi^+)B(\phi\to K^+K^-)=(3.6\pm0.9)\%$ and
$B(\Lambda_c^+\to pK^-\pi^+)=(5.0\pm1.3)\%$ \cite{pdg}, except for the $D_s^+$
data of Ref.~\cite{tasso}.
In the latter case, for consistency, we adopt the value
$B(D_s^+\to\phi\pi^+)B(\phi\to K^+K^-)=0.13\pm0.03\pm0.04$ from
Ref.~\cite{tasso} itself.

The differential cross sections $d\sigma/dx$ for the $D^0$, $D^+$, $D_s^+$,
and $\Lambda_c^+$ hadrons measured by CLEO \cite{cleo04,cleo0,cleo88}
(circles), HRS \cite{hrs88,hrs85} (squares), and TASSO \cite{tasso} (diamonds)
are confronted with our LO (dashed lines) and NLO (solid lines) predictions in
Figs.~\ref{fig:xs1}(a)--(d), respectively.

Let us first concentrate on our NLO predictions.
As for the $D^0$, $D^+$, and $D_s^+$ mesons, we observe that our NLO
predictions generally lead to a satisfactory description of the experimental
data, both in normalization and shape.
In particular, the maxima of the measured $x$ distributions are approximately
reproduced.
However, in the case of the $\Lambda_c^+$ baryon, the predicted $x$
distribution appears to be too hard, its peak being set off by approximately
$+0.2$ relative to the one shown by the experimental data.
In particular, the data points at 0.55 and 0.65 are poorly described by the NLO
prediction.
Although the $\Lambda_c^+$ baryon is 22\% heavier than the $D^0$ and $D^+$
mesons, and 16\% heavier than the $D_s^+$ meson, mass effects are unlikely to
be responsible for this disagreement, since $\sqrt s=10.55$~GeV is
sufficiently far above the charm threshold.

Let us now include the LO predictions in our considerations.
The CLEO data \cite{cleo04,cleo0,cleo88}, which are most precise, clearly
favor the NLO predictions, while the LO predictions are too large at small
values of $x$ and too small in the peak region.
Unfortunately, the HRS \cite{hrs88,hrs85} and TASSO \cite{tasso} data do not
reach the small-$x$ regime, where the LO and NLO predictions depart from each
other, and their errors are too large in order to support this observation.

Actually, the CLEO data \cite{cleo04,cleo0,cleo88} are considerably more
precise than the OPAL data \cite{opal1}, which we fitted to, and it would be
desirable to also include them in ours fits.
However, we refrain from doing so for the time being because their high
precision would make it necessary to properly treat finite-$m_Q$ effects,
which are neglected altogether in the theoretical formalism employed here.
The general-mass variable-flavor-number (GM-VFN) scheme \cite{acot}, which has
recently been extended to inclusive $X_Q$-hadron production in $\gamma\gamma$
\cite{ks1}, $ep$ \cite{ks2}, and $p\overline{p}$ \cite{kkss} collisions,
provides a rigorous theoretical framework that retains the full finite-$m_Q$
effects while preserving the indispensible virtues of the factorization
theorem \cite{col}, namely the universality and the DGLAP \cite{dglap} scaling
violations of the FF's entailing the resummation of dominant logarithmic
corrections.
A global analysis of experimental data on inclusive $X_c$-hadron production in
the GM-VFN scheme is left for future work.

\section{Conclusions}
\label{sec:five}

The OPAL Collaboration presented measurements of the fractional energy spectra
of inclusive $D^0$, $D^+$, $D_s^+$, and $\Lambda_c^+$ production in $Z$-boson
decays based on their entire LEP1 data sample \cite{opal1}.
Apart from the full cross sections, they also determined the contributions
arising from $Z\to b\bar b$ decays.
This enabled us to determine LO and NLO sets of FF's for these $X_c$ hadrons.

As in our previous analysis of $D^{*+}$ FF's \cite{bkk}, we worked in the
QCD-improved parton model implemented in the pure $\overline{\mathrm{MS}}$
renormalization and factorization scheme with $n_f=5$ massless quark flavors
(zero-mass variable-flavor-number scheme).
This scheme is particularly appropriate if the characteristic energy scale of
the considered production process, {\it i.e.}, the c.m.\ energy $\sqrt s$ in
the case of $e^+e^-$ annihilation and the transverse momentum $p_T$ of the
$X_c$ hadron in other scattering processes, is large compared to the
bottom-quark mass $m_b$.
Owing to the factorization theorem, the FF's defined in this scheme satisfy
two desirable properties:
(i) their scaling violations are ruled by the timelike DGLAP equations; and
(ii) they are universal.
Thus, this formalism is predictive and suitable for global data analyses.

We verified that the values of the branching and average momentum fractions of
the various $c,b\to X_c$ transitions evaluated at LO and NLO
using our FF's are in reasonable agreement with the corresponding results from
OPAL \cite{opal1} and other experiments \cite{pad,h1,aleph,delphi,gla}.

We tested the scaling violations of our FF's by comparing the fractional
energy spectra of inclusive $D^0$, $D^+$, $D_s^+$, and $\Lambda_c^+$
production measured in nonresonant $e^+e^-$ annihilation at $\sqrt s=10.55$~GeV
\cite{cleo04,cleo0,cleo88}, 29~GeV \cite{hrs88,hrs85}, and 34.7 \cite{tasso}
with our LO and NLO predictions to find reasonable agreement.
Since events of $X_c$-hadron production from $X_b$-hadron decay were excluded
from the data samples at $\sqrt s=10.55$~GeV, we obtained a clean test of our
charm-quark FF's.

It is important to bear in mind that the fit results for the input parameters
in Eqs.~(\ref{eq:peterson}) and (\ref{eq:standard}), including the value of
Peterson's $\epsilon$ parameter, are highly scheme dependent at NLO, and must
not be na\"\i vely compared without careful reference to the theoretical
framework which they refer to.

\bigskip
\centerline{\bf ACKNOWLEDGMENTS}
\smallskip\noindent
We thank A. J. Martin for providing to us the OPAL data \cite{opal1} in
numerical form.
B.A.K. thanks the Department of Energy's Institute for Nuclear Theory at the
University of Washington for its hospitality and the Department of Energy for
partial support during the completion of this work.
This work was supported in part by the Deutsche Forschungsgemeinschaft through
Grant No.\ KN~365/3-1, and by the Bundesministerium f\"ur Bildung und Forschung
through Grant No.\ 05~HT1GUA/4.

\newpage

\newpage
\begin{figure}[ht]
\begin{center}
\epsfig{figure=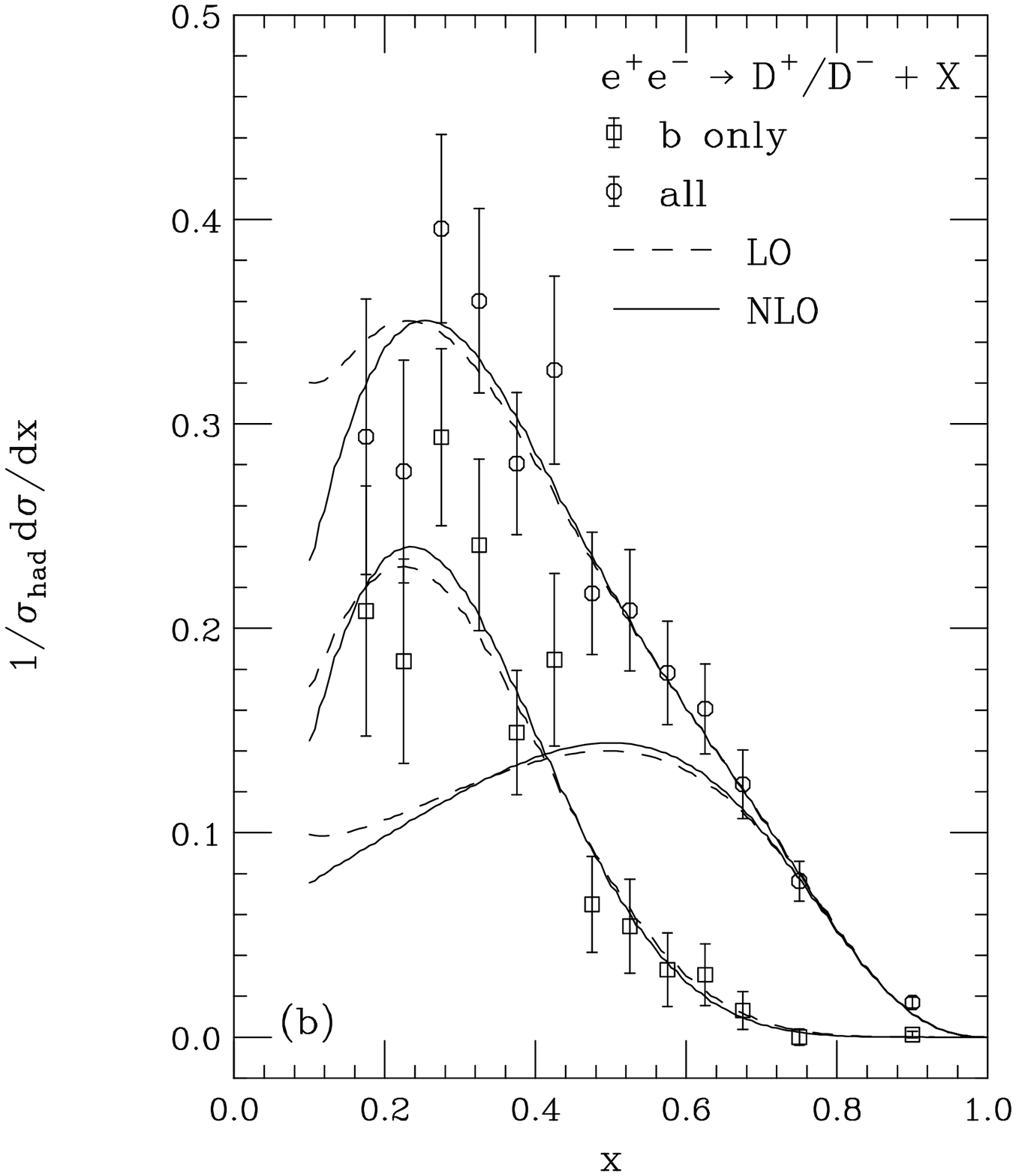,width=\textwidth}\\
Fig.~\ref{fig:xs} (continued).
\end{center}
\end{figure}

\newpage
\begin{figure}[ht]
\begin{center}
\epsfig{figure=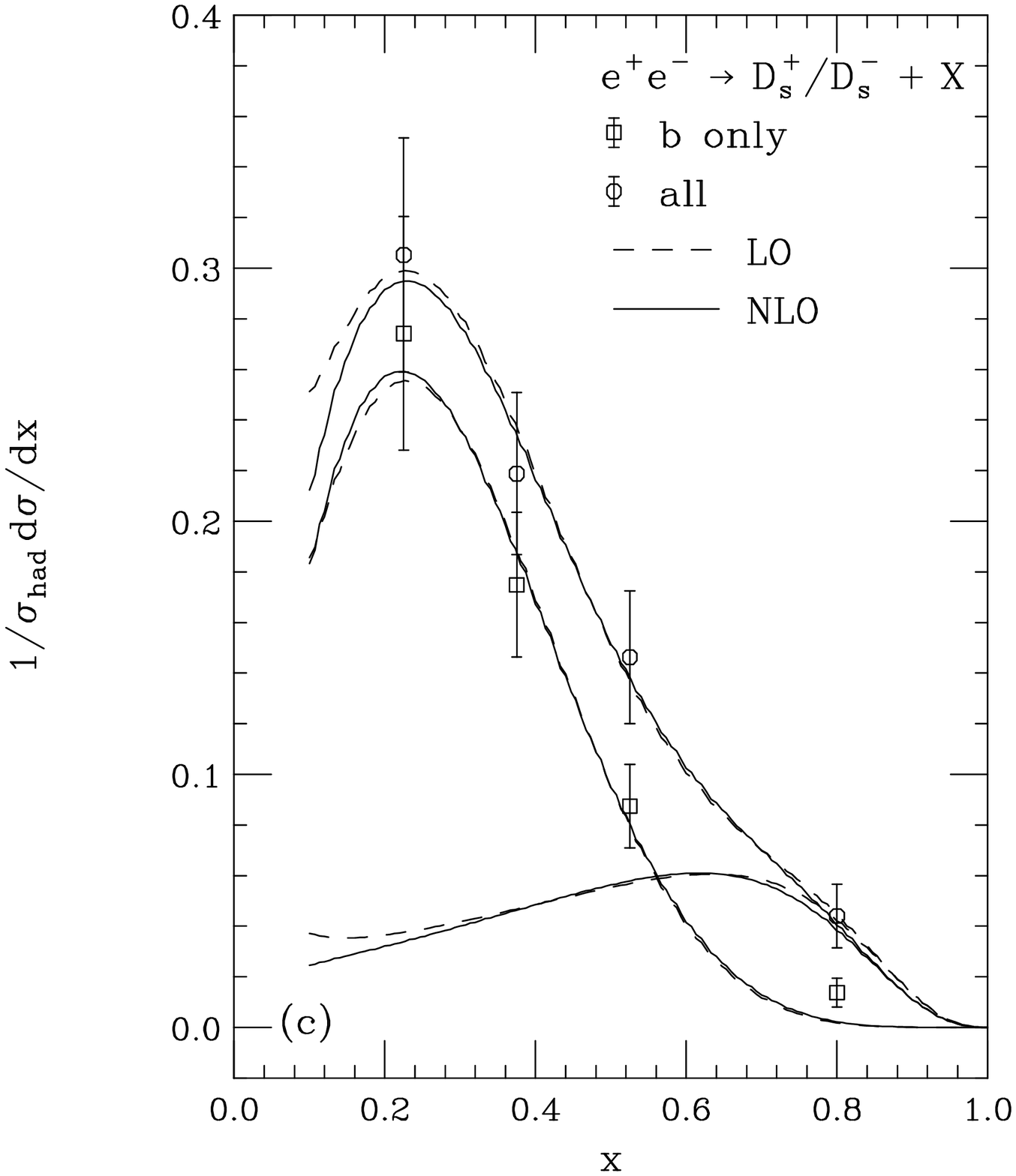,width=\textwidth}\\
Fig.~\ref{fig:xs} (continued).
\end{center}
\end{figure}

\newpage
\begin{figure}[ht]
\begin{center}
\epsfig{figure=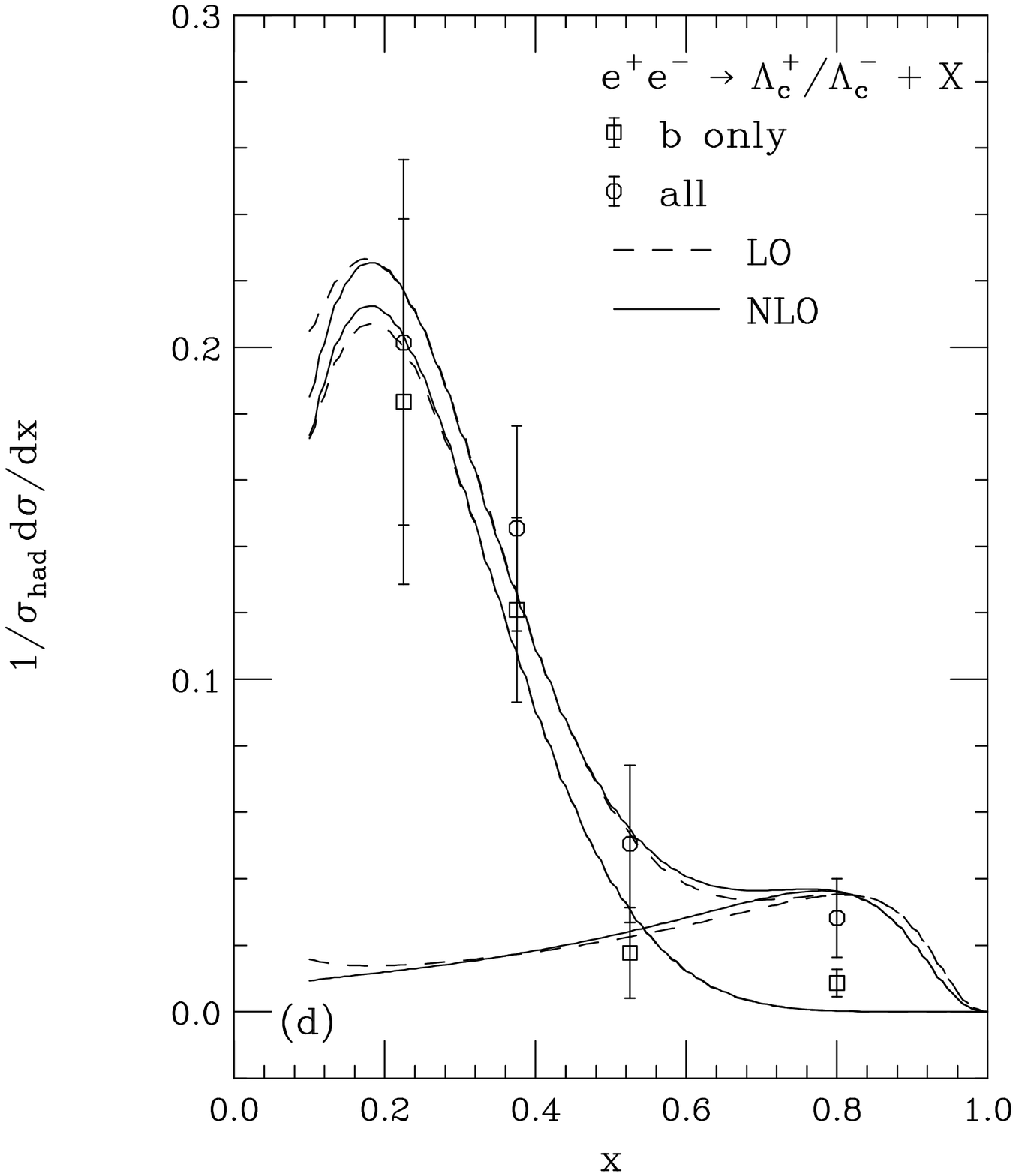,width=\textwidth}\\
Fig.~\ref{fig:xs} (continued).
\end{center}
\end{figure}

\newpage
\begin{figure}[ht]
\begin{center}
\epsfig{figure=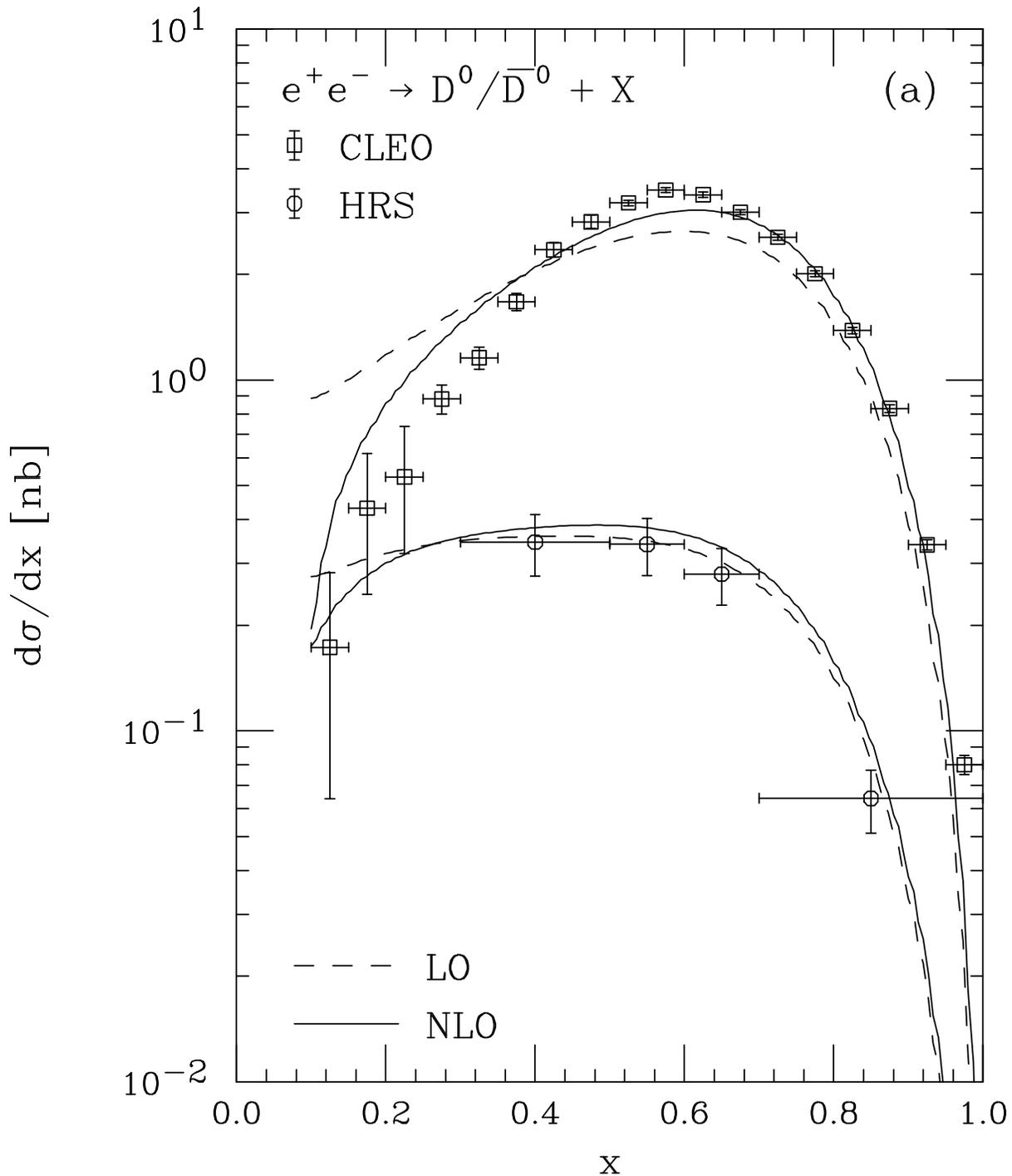,width=\textwidth}
\caption{The differential cross sections $d\sigma/dx$ (in nb) of inclusive (a)
$D^0/\overline{D}^0$, (b) $D^\pm$, (c) $D_s^\pm$, and (d) $\Lambda_c^\pm$
production in $e^+e^-$ annihilation at $\sqrt s=10.55$, 29, and $34.7$~GeV
evaluated at LO (dashed lines) and NLO (solid lines) with our respective FF
sets are compared with data from CLEO at CESR (squares), HRS at PEP (circles),
and TASSO at PETRA (diamonds), respectively.
In each case, the $X_c$ hadron and its charge-conjugate counterpart are summed
over.}
\label{fig:xs1}
\end{center}
\end{figure}

\newpage
\begin{figure}[ht]
\begin{center}
\epsfig{figure=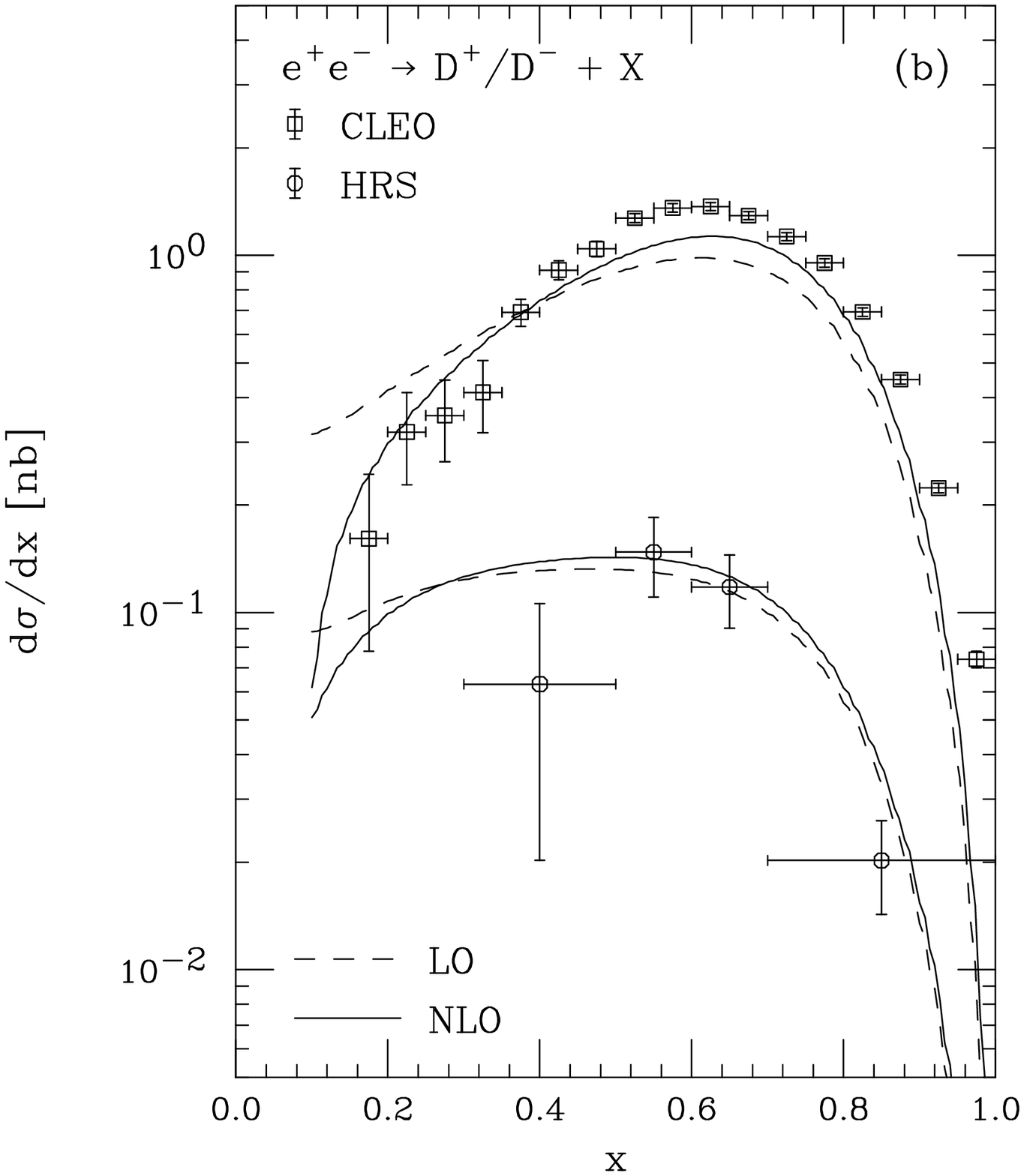,width=\textwidth}\\
Fig.~\ref{fig:xs1} (continued).
\end{center}
\end{figure}

\newpage
\begin{figure}[ht]
\begin{center}
\epsfig{figure=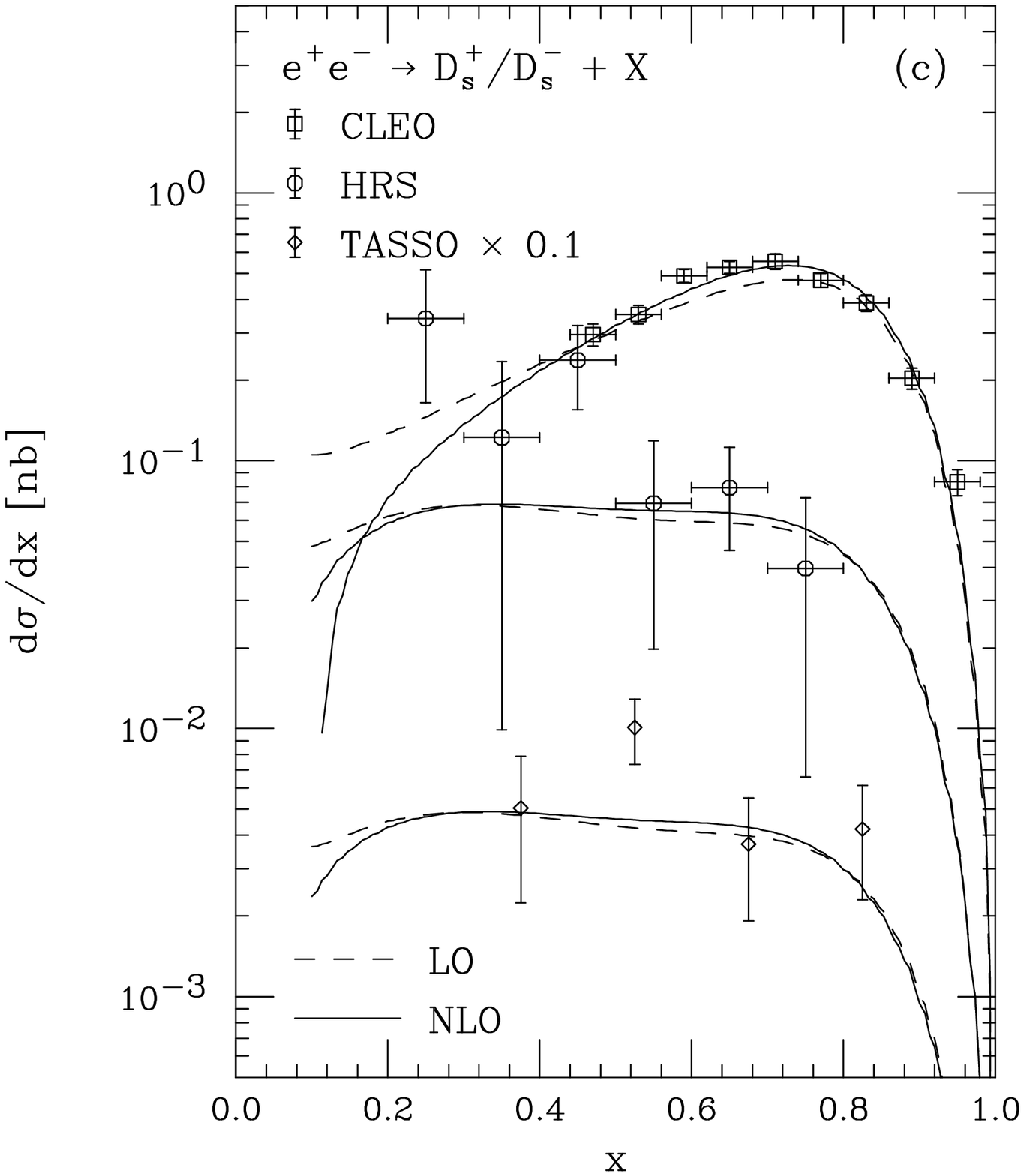,width=\textwidth}\\
Fig.~\ref{fig:xs1} (continued).
\end{center}
\end{figure}

\newpage
\begin{figure}[ht]
\begin{center}
\epsfig{figure=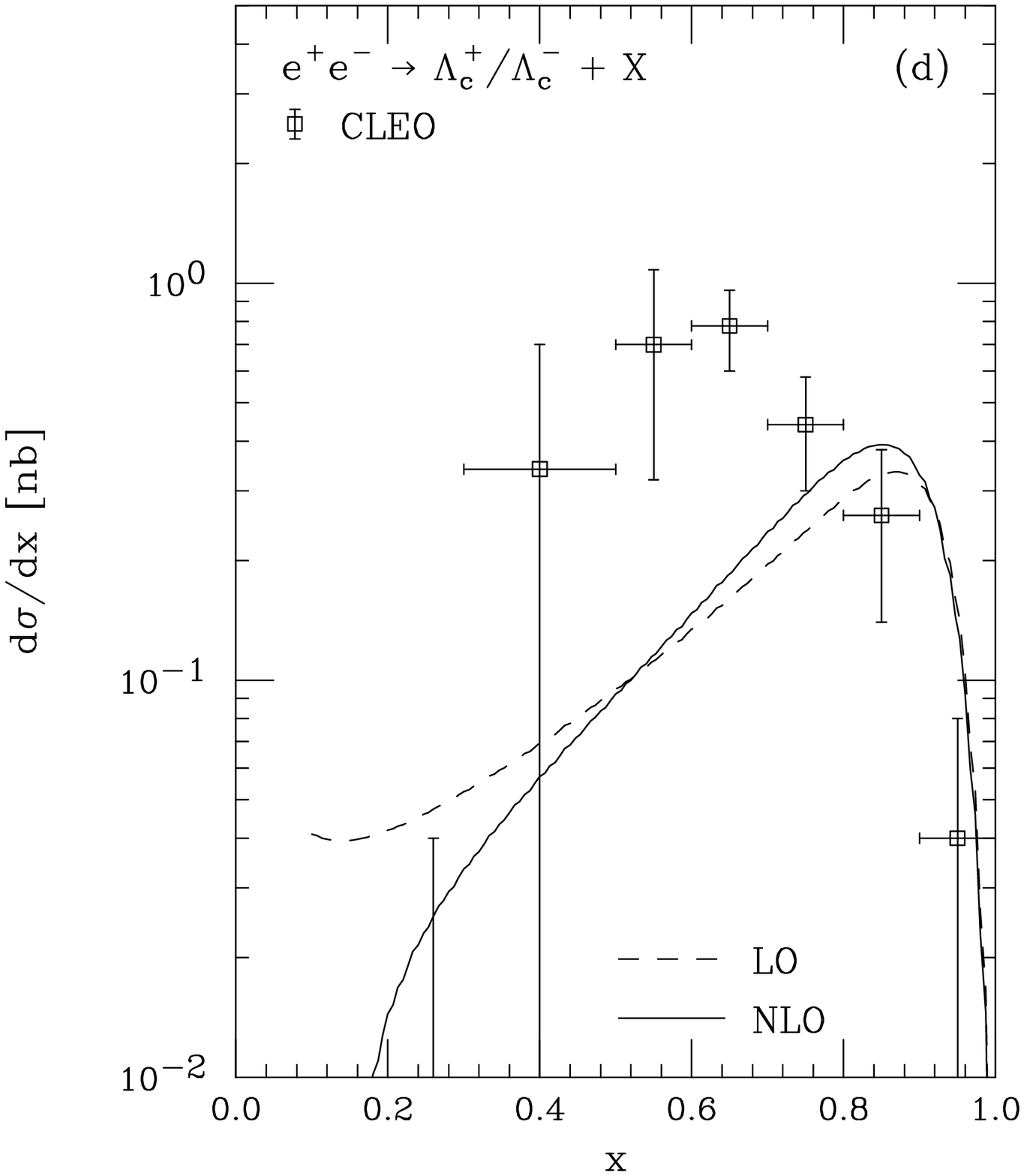,width=\textwidth}\\
Fig.~\ref{fig:xs1} (continued).
\end{center}
\end{figure}


\begin{thebibliography}{99}

\bibitem{zeus} ZEUS Collaboration, J. Breitweg {\it et al.},
Phys.\ Lett.\ B {\bf481}, 213 (2000).

\bibitem{pad} S. Padhi, in {\it Proceedings of the Ringberg Workshop on New
Trends in HERA Physics 2003}, edited by G. Grindhammer, B. A. Kniehl,
G. Kramer, and W. Ochs, (World Scientific, Singapore, 2004), p.~183.

\bibitem{h1} H1 Collaboration, A. Aktas {\it et al.},
Eur.\ Phys.\ J. {\bf38}, 447 (2005).
% H1 Collaboration, contribution to the {\it International
%Europhysics  Conference on High Energy Physics (EPS03)}, Aachen, Germany,
%July 17--23, 2003, Abstract 096.

\bibitem{cdf} CDF Collaboration, D. Acosta {\it et al.},
Phys.\ Rev.\ Lett.\ {\bf91}, 241804 (2003).

\bibitem{aleph} ALEPH Collaboration, R. Barate {\it et al.},
Eur.\ Phys.\ J. C {\bf16}, 597 (2000).

\bibitem{opal} OPAL Collaboration, K. Ackerstaff {\it et al.},
Z.\ Phys.\ C {\bf1}, 439 (1998).

\bibitem{bkk} J. Binnewies, B. A. Kniehl, and G. Kramer,
Phys.\ Rev.\ D {\bf58}, 014014 (1998).

\bibitem{opal1} OPAL Collaboration, G. Alexander {\it et al.},
Z.\ Phys.\ C {\bf72}, 1 (1996).

\bibitem{spira} B. A. Kniehl, M. Kr\"amer, G. Kramer, and M. Spira,
Phys.\ Lett.\ B {\bf356}, 539 (1995).

\bibitem{bkk1} J. Binnewies, B. A. Kniehl, and G. Kramer,
Z.\ Phys.\ C {\bf76}, 677 (1997).

\bibitem{dglap} V. N. Gribov and L. N. Lipatov,
Yad.\ Fiz.\ {\bf15}, 781 (1972)
[Sov.\ J. Nucl.\ Phys.\ {\bf15}, 438 (1972)];
G. Altarelli and G. Parisi,
Nucl.\ Phys.\ {\bf B126}, 298 (1977);
Yu.~L. Dokshitzer,
Zh.\ Eksp.\ Teor.\ Fiz.\ {\bf73}, 1216 (1977)
[Sov.\ Phys.\ JETP {\bf46}, 641 (1977)].

\bibitem{pet} C. Peterson, D. Schlatter, I. Schmitt, and P. M. Zerwas,
Phys.\ Rev.\ D {\bf27}, 105 (1983).

\bibitem{kks} B. A. Kniehl, G. Kramer, and M. Spira,
Z.\ Phys.\ C {\bf76}, 689 (1997).

\bibitem{bkk2} J. Binnewies, B. A. Kniehl, and G. Kramer,
Phys.\ Rev.\ D {\bf52}, 4947 (1995).

\bibitem{martin} A. J. Martin, private communication.

\bibitem{pdg} Particle Data Group, S. Eidelman {\it et al.},
Phys.\ Lett.\ B {\bf592}, 1 (2004).

\bibitem{delphi} DELPHI Collaboration, P. Abreu {\it et al.},
Eur.\ Phys.\ J. C {\bf12}, 225 (2000).

\bibitem{gla} L. Gladilin,
Report No.\ hep-ex/9912064 (unpublished).

\bibitem{cleo04} CLEO Collaboration, M. Artuso {\it et al.},
Phys.\ Rev.\ D {\bf70}, 112001 (2004).

\bibitem{cleo0} CLEO Collaboration, R. A. Briere {\it et al.},
Phys.\ Rev.\ D {\bf62}, 072003 (2000).

\bibitem{cleo88} CLEO Collaboration, D. Bortoletto {\it et al.},
Phys.\ Rev.\ D {\bf37}, 1719 (1988).

\bibitem{hrs88} HRS Collaboration, P. Baringer {\it et al.},
Phys.\ Lett.\ B {\bf206}, 551 (1988).

\bibitem{hrs85} HRS Collaboration, M. Derrick {\it et al.},
Phys.\ Rev.\ Lett.\ {\bf54}, 2568 (1985).

\bibitem{tasso} TASSO Collaboration, M. Althoff {\it et al.},
Phys.\ Lett.\ {\bf136B}, 130 (1984).

\bibitem{acot} M. A. G. Aivazis, J. C. Collins, F.I. Olness, and W.-K. Tung,
Phys.\ Rev.\ D {\bf50}, 3102 (1994).

\bibitem{ks1} G. Kramer and H. Spiesberger,
Eur.\ Phys.\ J.\ C {\bf22}, 289 (2001);
{\it ibid.} {\bf28}, 495 (2003).

\bibitem{ks2} G. Kramer and H. Spiesberger,
Eur.\ Phys.\ J.\ C {\bf38}, 309 (2004).

\bibitem{kkss} B. A. Kniehl, G. Kramer, I. Schienbein, and H. Spiesberger,
Phys.\ Rev.\ D {\bf71}, 014018 (2005);
Report No.\ DESY 05-030, MZ-TH/05-04, hep-ph/0502194, Eur.\ Phys.\ J.\ C
(in press).

\bibitem{col} J. C. Collins,
Phys.\ Rev.\ D {\bf58}, 094002 (1998).

\end{thebibliography}
\end{document}